\documentclass[journal=jpclcd,manuscript=letter]{achemso}
\usepackage{amsmath,color}
\usepackage{graphicx}
\usepackage{amssymb}
\usepackage{caption}
\usepackage{subcaption}
\usepackage{rotating}
\usepackage[version=3]{mhchem}

\author{Sandeep K. Jain}
\affiliation[Universiteit Utrecht]
{Institute for Theoretical Physics, Universiteit Utrecht, Leuvenlaan 4, 3584 CE Utrecht, The Netherlands}
\email{S.K.Jain@uu.nl}
\author{Vladimir  Juri\v ci\'c}
\affiliation{Institute for Theoretical Physics, Universiteit Utrecht, Leuvenlaan 4, 3584 CE Utrecht, The Netherlands}
\email{V.Juricic@uu.nl}
\author{Gerard T. Barkema}
\affiliation[Institute for Theoretical Physics,Universiteit Utrecht]
{Institute for Theoretical Physics, Universiteit Utrecht, Leuvenlaan 4, 3584 CE Utrecht, The Netherlands}
\alsoaffiliation[Universiteit Leiden]
{Institut-Lorentz, Universiteit Leiden, Niels Bohrweg 2, 2333 CA Leiden, The Netherlands}
\title{Probing crystallinity of graphene samples via the vibrational density of states}

\begin{document}

\begin{abstract}
The purity of graphene samples is of crucial importance for their
experimental and practical use. In this
regard, the detection of the defects
is of direct relevance. Here, we show that structural defects in graphene samples give rise to clear signals
in the vibrational density of states (VDOS) at the specific peaks at high and low frequencies.
These can be used as an independent probe of the defect density. In particular,
we consider grain boundaries made of pentagon-heptagon pairs, and
show that they lead to a shift of the characteristic vibrational D
mode towards higher frequency; this distinguishes these line defects
from the Stone-Wales point defects, which do not lead to such a shift.
Our findings may be instrumental for the detection of structural lattice
defects using experimental techniques that can directly measure VDOS,
such as inelastic electron tunneling and inelastic neutron spectroscopy.
\end{abstract}

Since it has been experimentally isolated, graphene
has attracted enormous attention due to its unusual electronic and
mechanical properties \cite{RMP-CastroNeto2009}. The quality of
the crystalline samples is very important for the observation of
the hallmark features of graphene, such as ballistic conductivity \cite{andrei-natnanotech-2008,bolotin-2008},
 as well as for its mechanical and chemical properties, e.g.,  its permeability  \cite{nair-science-2012}.
Large graphene samples produced, for instance, by chemical vapor-deposition (CVD), exfoliation
or epitaxial growth on metal and SiC substrates are typically polycrystalline and
thus contain intrinsic lattice defects, such as grain boundaries
\cite{yazyev-review-2014,rasool-nanoletters-2014,tison-nanoletters-2014},
dislocations and Stone-Wales (SW) defects, as well as extrinsic defects,
e.g., adatoms \cite{Dresselhaus-review}.  Detection of the lattice defects
is therefore of both fundamental and practical relevance, since they
are inevitably present in the graphene samples and can significantly
alter the graphene's chemical and physical properties \cite{banhart-review-2011}. In fact, the defects may not only be
detrimental for the properties of graphene, but may also be interesting
in their own right, as they may lead to some new effects, not present
otherwise \cite{Lu-review-2013,Liu-review-2015,Guryel2013}, and are also important for graphene nano-devices \cite{Vicarelli2015}.

\begin{figure}
\centering
\includegraphics[width=1.0\textwidth]{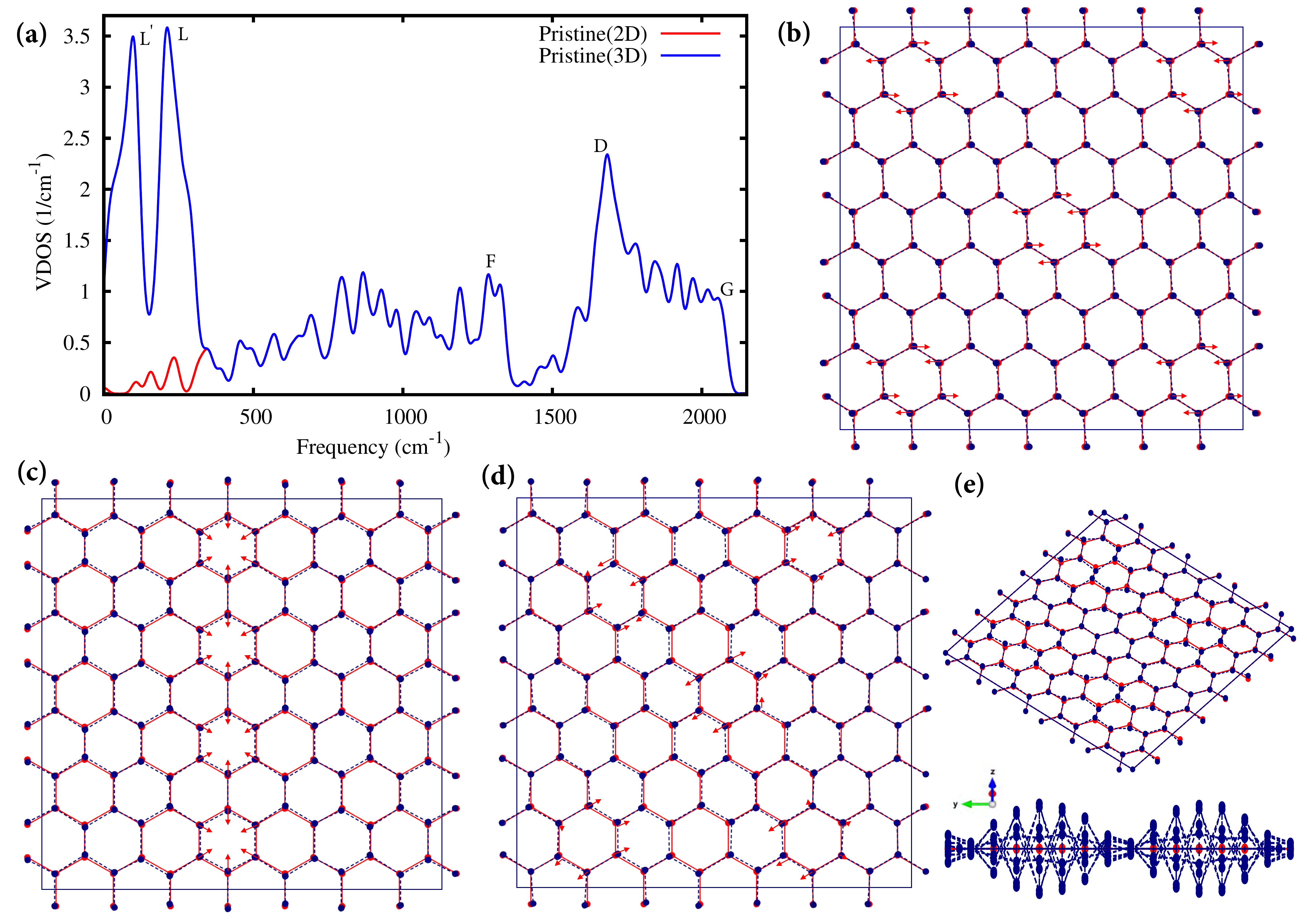}
\caption{VDOS and the profile of the displacements of the
prominent modes in pristine graphene. (a) VDOS of both flat and buckled
pristine graphene. (b) Vibrational mode corresponding to Raman active
G mode at 2080 cm$^{-1}$. (c) D mode at 1660 cm$^{-1}$. (d) F mode at
1280 cm$^{-1}$. (e) Out-of-plane L mode at 210 cm$^{-1}$: displacements in the flat graphene's ($x-y$) plane (top) and in the side $y-z$ plane (bottom).
}
\label{Fig-1}
\end{figure}

Structural defects are especially prominent in this regard
\cite{banhart-review-2011}. In particular, graphene is a unique
two-dimensional crystalline membrane that hosts lattice defects
arising due to the flexibility of the carbon atoms in hybridization. As a
result, polygons different from hexagons can appear in the lattice
structure.  Energetically favorable point-like defects of this type
include the SW defect obtained when four hexagons are transformed
by a bond transposition of $90^\circ$ into two pentagon-heptagon
pairs, thereby conserving the number of the atoms.
They can be formed thermally in pristine graphene, but
have a formation energy of $\sim 5$ eV, and thus pristine graphene may
host only a few of them. On the other hand, such defects and similar
ones can be frozen in during annealing process and it is therefore not
surprising that they have been experimentally observed \cite{nature-2004,
meyer-2008,meyer-2011}. Moreover, there have been proposals for their
controllable production in  graphene \cite{Lu-review-2013}.

Techniques to characterize the crystal structure of graphene include
direct local ones, such as transmission electron microscopy (TEM),
scanning tunneling microscopy (STM) \cite{andrei-review-2012} and
atomic force microscopy (AFM) \cite{ingmar-2014,dedkov-2014},
as well as the indirect ones, among which Raman spectroscopy
\cite{Kudin2008,ferrari-review-2013,Matz2015}, X-ray absorption spectroscopy \cite{Lee2010,DeJesus2013,Mohr2007}, inelastic electron tunneling spectroscopy (IETS)
\cite{vitali-prb-2004,cervenka-prb-2010,palsgaard-prb-2015}, and neutron
scattering \cite{rols-2000,cavallari-neutron-scattering-2014}.
Although widely used, Raman spectroscopy is limited by selection rules to only
a certain number of Raman-active vibrational modes, which include the
so-called G and 2D peaks located at 1580 cm$^{-1}$ and 2680 cm$^{-1}$,
respectively \cite{ferrari-prl-2006}, originating from the G and D
phonon modes \cite{dresselhaus-book}. In the presence of disorder,
due to the breaking of the lattice symmetry,
the D mode at 1340 cm$^{-1}$, Raman inactive in pristine graphene, becomes active.
However, little is known about the
specific experimentally observable signatures of the structural defects,
such as point and line defects, in the vibrational spectrum. This is an important problem especially in light of the recent
mapping of the entire vibrational spectrum of graphene by IETS \cite{naterrer-prl-2015}, and reported signature vibrational bands in CVD graphene with defects \citep{Matz2015}.

\begin{figure}
\includegraphics[width=1.0\textwidth]{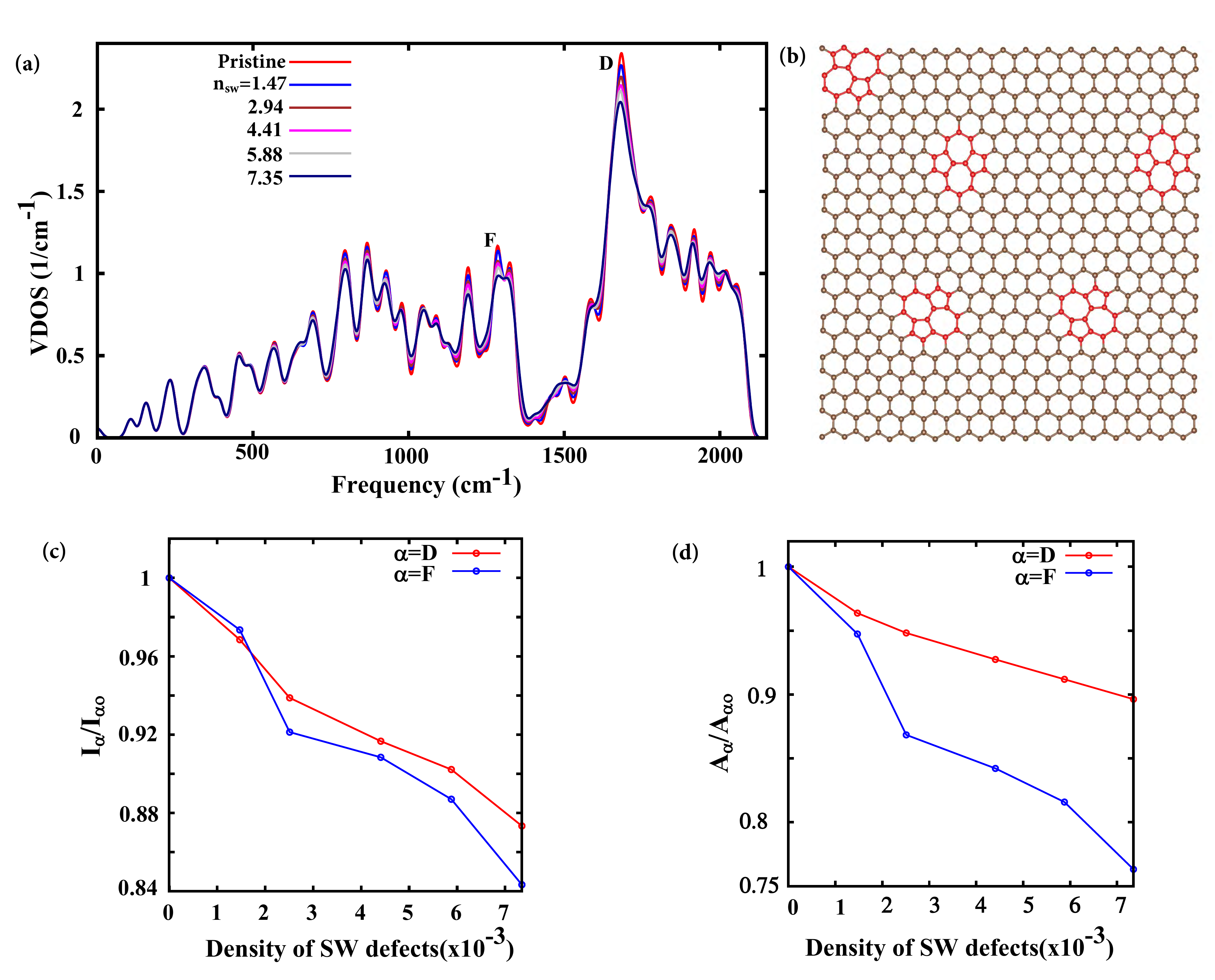}
\caption{Structure of a flat graphene sample having multiple point SW defects and corresponding VDOS.  (a) VDOS of flat graphene samples with different
densities of SW defects $n_{SW}$($\times 10^{-3}$). (b) The flat graphene
sample with five SW defects. (c) Relative decrease in the intensity of
D and F peaks at different defect densities. (d) Relative decrease in the number of modes of D and F bands at different defect densities in the range of frequencies in which VDOS is greater than 1.5/cm$^{-1}$ and $1.0$/cm$^{-1}$, respectively.}  
\label{Fig-2}
\end{figure}

\begin{figure}
\includegraphics[ width=1.0\textwidth]{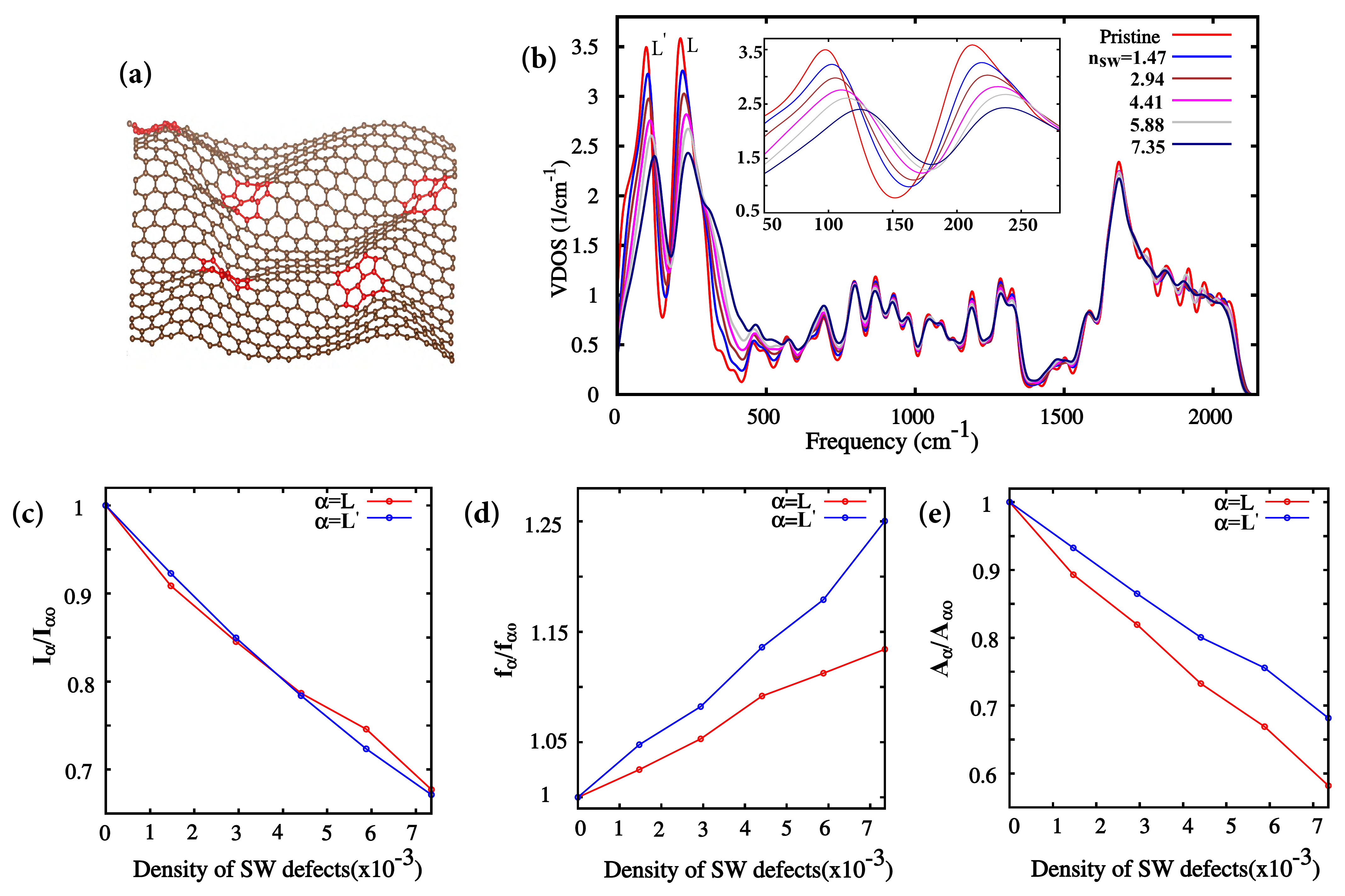}
\caption{Structure of a buckled graphene sample having multiple point SW defects and corresponding VDOS. (a) The lattice structure of a buckled graphene
sample with five SW defects. (b) VDOS of buckled graphene samples
with different densities of SW defects $n_{SW}$($\times10^{-3}$).
Inset: Low-frequency peaks in VDOS zoomed in. (c)
Relative decrease in the intensity of L and L' modes at different
$n_{SW}$. (d) Relative increase (blue shift) in the frequency of L and
L' modes at different $n_{SW}$. (e) Relative decrease in the number of modes of L and L' bands
at different defect densities in the range of frequencies in which VDOS is greater than 1.9/cm$^{-1}$ and $1.5$/cm$^{-1}$, respectively.}  
\label{Fig-3}
\end{figure}

In this Letter, we show that the nature and density of structural point and line
defects in graphene samples can be characterized by the specific and distinct features
in the VDOS. These features are directly detectable in IETS and neutron scattering, which are not limited by selection
rules, as opposed to the Raman spectroscopy, and can thus probe the
entire vibrational spectrum \cite{naterrer-prl-2015}.
Specifically, using a recently developed
effective semiempirical elastic potential \cite{Sandeep-2015} we show
that the presence of the point-like SW defects in pristine flat graphene gives
rise to a simultaneous decrease in the VDOS of the peaks corresponding
to high-frequency D and F modes [Figs.\ \ref{Fig-1}(a) and \ref{Fig-2}].
More importantly,
the graphene membrane has a natural
tendency to buckle, and as a result new low-energy vibrational
states appear, with a particularly pronounced L and L' peaks [Fig.\
\ref{Fig-1}(a)]. When the SW defects are introduced, the intensities of
these characteristic modes simultaneously decrease, and the peak positions
shift toward higher values of frequency (blue shift), see Fig.\ \ref{Fig-3}.
On the other hand, we
find that line defects give rise to a blue shift of the
D peak in conjunction with the decrease of its intensity. Substrate plays an important role in the production of graphene samples, \cite{Wang2008,Wood2011} and we therefore show that the decrease in the VDOS of low-frequency modes  without any shift signals its presence.

To calculate the vibrational spectrum of graphene, we use a
recently developed effective semiempirical elastic potential \cite{Sandeep-2015}
\begin{align}\label{eff-potential}
E&=\frac{3}{16}\frac{\alpha}{d^2}\sum_{i,j}(r_{ij}^2-d^2)^2
  +\frac{3}{8}\beta d^2\sum_{j,i,k}\left(\theta_{jik}-\frac{2\pi}{3}\right)^2
  +\gamma\sum_{i,jkl}r_{i,jkl}^2.
\end{align}
Here,
$r_{ij}$ is the distance between two bonded atoms,
$\theta_{jik}$ is the angle between the two bonds connecting atom $i$
to atoms $j$ and $k$, and $r_{i,jkl}$ is the distance between atom $i$
and the plane through the three atoms $j$, $k$ and $l$ connected to
atom $i$.
The parameter $\alpha=26.060$~eV/\AA$^{2}$ controls bond-stretching
and is fitted to the bulk modulus, $\beta=5.511$~eV/\AA$^{2}$ controls bond-shearing
and is fitted to the shear modulus,
$\gamma=0.517$~eV/\AA$^{2}$ describes the stability of the graphene sheet against buckling,
and $d = 1.420$ \AA~is the ideal bond length for graphene.
The parameters in the potential (eq. 1)
are obtained by fitting to density-functional theory (DFT) calculations \cite{Sandeep-2015}.
Additionally, the effect of the substrate to the buckled graphene sample is described
by an extra harmonic term
\begin{equation}\label{eq-substrate}
E_S=K\sum_{i}z_i^2,
\end{equation}
where $K$ is the effective elastic constant for the graphene-substrate
interface, and $z_i$ is the distance of atom $i$ from the graphene plane.

The vibrational spectrum is obtained from the above potential. The VDOS
represents the number of modes at a certain frequency, and the total area
under the VDOS  gives the total number of vibrational modes, which is
$2N$ for a flat and $3N$ for a buckled graphene sheet, with $N$ the
total number of atoms in the sample.  In our plots, the VDOS is convoluted
with a gaussian function with a width of $\sigma=14$ cm$^{-1}$, and
$N=680$; this is much larger than system sizes
in previous {\it ab initio} studies of pristine graphene and graphene with defects
\cite{ma-2009,koskinen-2009,mazzamuto-2011,shirodkar-2012}. Furthermore, we have computed the relative decrease of the
number of modes corresponding to the characteristic bands  as a function of the defect density.
We have performed this computation by counting the number of modes with VDOS greater than a certain fixed value selected by the symmetry of the characteristic peak, see Figs.\ \ref{Fig-2}(d) and \ref{Fig-3}(e).

{\it Pristine flat and buckled graphene sheet:-} We first calculate the
VDOS for a flat pristine graphene sample, which serves as the
reference spectrum in the following. The plot displayed in Fig.\
\ref{Fig-1}(a) shows the characteristic peaks that correspond to D
and G vibrational bands, of which only the latter is Raman active
\cite{dresselhaus-book}. The corresponding modes, shown
in Figs.\ \ref{Fig-1}(b) and \ref{Fig-1}(c) transform under E$_{2g}$ and A$_{1g}$ representations of the graphene's D$_{6h}$ point group symmetry, respectively \cite{mappelli-1999}.
Notice that the hallmark feature
of the monolayer graphene, the Raman-active G mode, which is at the
maximum frequency in the vibrational spectrum, see Fig.\ \ref{Fig-1},
is positioned at $f_{G}\simeq2080$ cm$^{-1}$, and therefore offset
by about 25\% as compared to the experimentally measured value of
$\sim1580$ cm$^{-1}$. The D mode at $f_D\simeq1660$ cm$^{-1}$
is also shifted by about 25\% compared to the experimentally measured
$1340$ cm$^{-1}$.  Deviations of this magnitude are expected since the parameters of the
effective potential are obtained from the DFT  \cite{Sandeep-2015} with similar deviations. However, the form of the displacements
of these modes, shown in Fig.\ \ref{Fig-1}(b) and \ref{Fig-1}(c), allows
us to identify them as G and D modes, respectively.  Furthermore,
the vibrational spectrum of graphene features the F mode,  shown in Fig.\
\ref{Fig-1}(d), at $f_F\simeq1280$ cm$^{-1}$. A flat graphene sheet has a
natural tendency to buckle, and this leads to the appearance of additional
soft out-of-plane phonon modes in the range of frequencies approximately
up to $f_{3D}\sim 300$ cm$^{-1}$, which is of the order of  the energy
scale corresponding to the parameter $\gamma$ describing the buckling
in the effective potential (eq. 1), $f^\gamma\sim(1/2\pi
c)\sqrt{\gamma/m_C}\simeq150$ cm$^{-1}$, with $m_C\simeq2\times10^{-26}$
kg as the atomic mass of carbon, and $c=3\times10^8$ m/s the velocity
of light. The L and L' modes, the former being a B$_{2g}$ mode [Fig.\
\ref{Fig-1}(e)], at frequency $f_L\simeq 210$ cm$^{-1}$ and  $f_{L'}\simeq
100$ cm$^{-1}$, respectively, are especially prominent and, as we show,
can be used to probe the point defects in the buckled graphene samples.

{\it Point SW defects in flat and buckled graphene samples:-} We now study
the VDOS in flat graphene with point-like SW defects , see Fig.\ \ref{Fig-2}(b). The
obtained VDOS is displayed in Fig.\ \ref{Fig-2}(a). We first observe
that significant changes in the VDOS occur in the high-frequency region, at
frequencies above 500 cm$^{-1}$. As the density of the defects increases, the
VDOS at the peaks decreases. On the other hand, the VDOS at the minima
increases, as a consequence of the conservation of the total number of
the vibrational modes. Notice in particular that the height of both D
and F peaks {\it simultaneously} decreases as more and more defects are added to
the sample, see Fig.\ \ref{Fig-2}(c). Particularly, for the highest defect
concentration considered $n_{SW}\simeq0.7\%$, the VDOS decreases by about
$12\%$ and $15\%$ for the D and F mode, respectively. This simultaneous
decrease of the two peaks in VDOS represents a hallmark feature of the
presence of point defects in the flat graphene sheet,
and is certainly experimentally observable. Furthermore, we have found
the relative decrease in the number of modes for D and F band of $\sim 10\%$ and $\sim 25\%$, respectively, see Fig.\ \ref{Fig-2}(d).

The SW defects in the buckled graphene sheet [Fig.\ \ref{Fig-3}(a)]
have a drastic effect on the low-frequency L and L' vibrational
bands. Their presence gives rise to the {\it simultaneous} decrease of
the corresponding peaks in the VDOS, together with the increase of
the mode corresponding to the minimum between the two maxima in VDOS,
as shown in Fig.\ \ref{Fig-3}(b). The decrease in VDOS of the two peaks
is proportional to the defect concentration [Fig.\ \ref{Fig-3}(c)], and it
appears to be significant. For instance, for the defect concentration
of $n_{SW}\simeq0.7$\% it is of the order of 30\%. Furthermore, this
decrease occurs in conjunction with a systematic blue shift of the
maximum of the two modes as the density of the defects increases, see
Fig.\ \ref{Fig-3}(d). In particular, for the defect density $n_{SW}\sim0.7$
\%, it is of the order of 25\% and 15\% for L' and L mode, respectively.
Finally, we find a significant relative decrease in the number of modes as a function of
the defect density, which is $\sim 40\%$ and $\sim 30\%$ for the L and L' bands, respectively, see Fig.\ \ref{Fig-3}(e).

\begin{figure}
\includegraphics[width=0.70\textwidth, height=18cm]{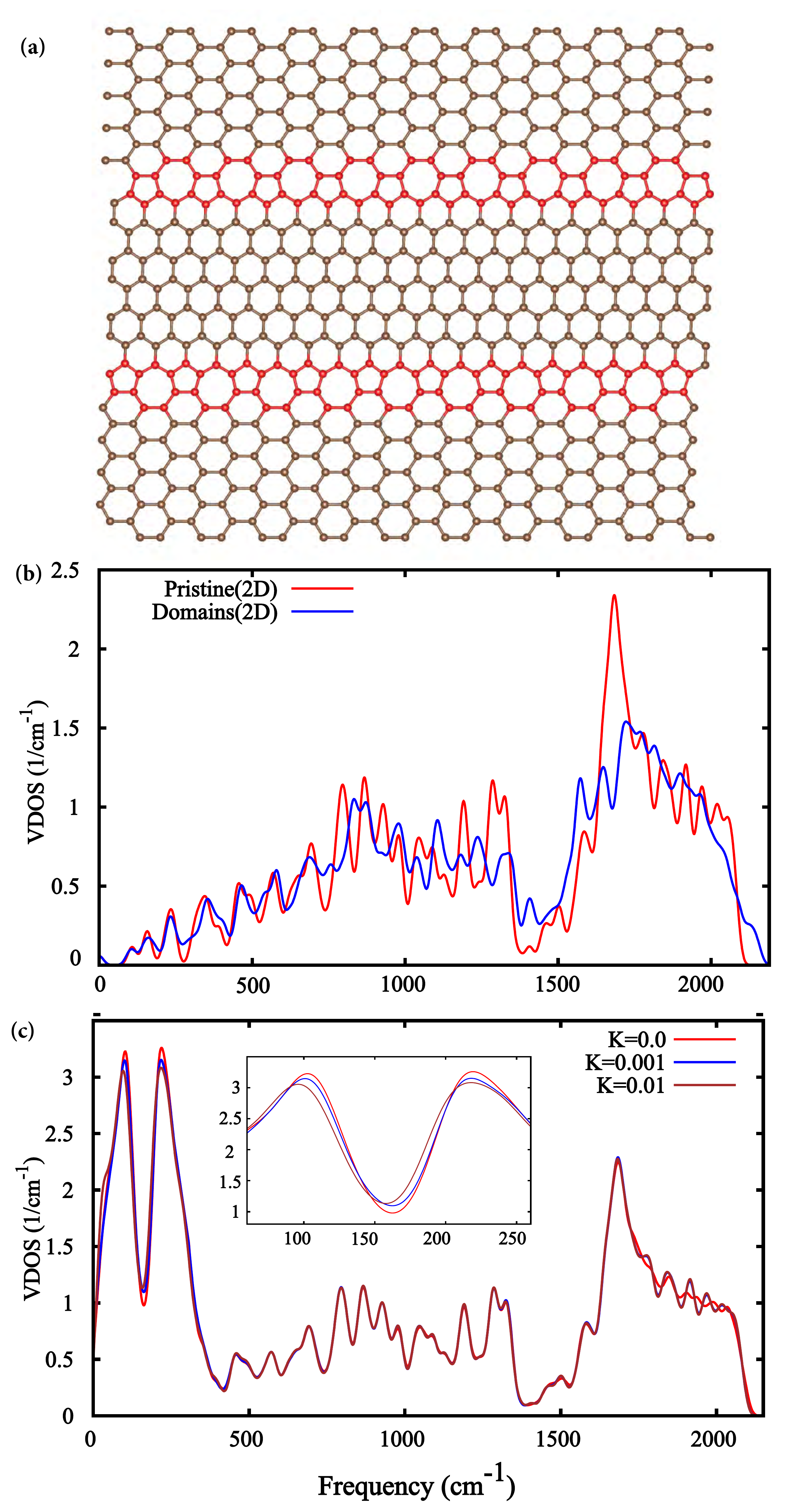}
\caption{Signature of the line defect and substrate on in the VDOS. (a) A graphene sample with two differently
oriented domains, with angular mismatch of $30^{o}$ separated by a
straight line of alternating pentagon and heptagon rings. (b) Comparison of the VDOS of
pristine graphene and graphene with grain boundaries. (c) VDOS of a graphene
sample having one SW defect interfaced by a substrate, with the confining
potential given by Eq.\ (eq. 2), and with the parameter $K$
in eV/{\AA}$^2$. Inset: Low-frequency part of VDOS with the prominent L and L' peaks. }
\label{Fig-4}
\end{figure}

{\it Signatures of the domains and the substrate in the VDOS:-} We now
turn to the effects of the grain boundary [Fig.\ \ref{Fig-4}(a)] to the
vibrational modes in flat graphene. As can be seen in Fig.\ \ref{Fig-4}(b),
the most prominent features in the VDOS are visible in the high-frequency
region. In particular, the intensity of both the F and D bands decreases
by approximately 35\%, followed by the simultaneous blue shift of both
bands by about  2\%. These effects should be contrasted to the behavior
of VDOS in the presence of the single point-like SW defects, where no such a shift
occurs either in confined two-dimensional geometry or in buckled samples. This blue
shift may be attributed to the fact that due to the presence of the
defects, the atomic bonds become shorter and therefore stiffer at the
position of the defects. The corresponding modes thus become shifted to
higher frequencies as compared to the defect-free sample. Notice that this
effect is negligible  in the case of the confined graphene sample because
the relative change of the parameter $\alpha$ in the effective potential
(eq. 1) is very small, less than $1\%$. In a buckled sample,
on the other hand, due to the fact that the out-of-plane modes are soft,
this change is much larger, $\sim 20\%$, and therefore leads to the
obtained more pronounced effect, see the inset of Fig.\ \ref{Fig-3}(b).
Finally, we consider the effect of the substrate described by the term
(eq. 2) in the effective potential. We plot VDOS for several
values of the parameter $K$ in Fig.\ \ref{Fig-4}(c), and observe that the
intensity of the low-frequency L and L' peaks decreases by about $5\%$
[see inset of Fig.\ \ref{Fig-4}(c)], without any shift in the position,
which is yet different from point defects where such a decrease is followed
by the blue shift of the peaks. Notice that this suppression of the intensity of the L and L' peaks due to the confinement by the substrate is in qualitative agreement with
the experimentally observed low-intensity out-of-plane phonon modes in the backgated graphene sample placed on a substrate \cite{naterrer-prl-2015}. Although the harmonic confinement potential (eq. 2) is certainly a crude description of the real interaction, it may be improved to also accurately capture the quantitative features of the experimentally observed VDOS spectrum. 

To conclude, we have shown that the VDOS
can be used as a tool to detect the presence of point and line defects in graphene sheets.
The confined two-dimensional graphene sample in the presence of the defects shows clear features
in the high-frequency VDOS, while the most pronounced effects of the
buckling appear at low-frequencies.
Given the recent measurement of the entire vibrational spectrum of backgated graphene placed on a substrate using IETS \cite{naterrer-prl-2015},
we hope that our findings will stimulate  further
experiments to probe the structural defects in this material.
Our results can also be used to calculate
the corresponding Raman response using different models \cite{rao-science-1997,basko-prb-2008}.
We would like to point out that our findings may also be applicable to other two-dimensional
materials that could be described by a semiempirical potential of the form
(eq. 1), such as nanoporous carbon \cite{bourgeois-1997,
smith-2009}. We hope that these results will stimulate further studies of vibrational properties of other carbon-based nanomaterials such as carbon-nanotubes \cite{Pekker2011,Xin2012}, graphene nano-ribbons, \cite{OuYang2008} and functionalized graphene \cite{Mattson2014}.

\begin{acknowledgement}
We acknowledge the support by FOM-SHELL-CSER program (12CSER049). This
work is part of the research program of the Foundation for Fundamental
Research of Matter (FOM), which is part of the Netherlands Organisation
for Scientific Research (NWO). We would like to thank Ingmar Swart and Rembert Duine for useful discussions.
\end{acknowledgement}

\bibliography{VDOS}

\end{document}